\documentclass[10pt,letterpaper,DIV=16,parskip=half,twoside]{scrartcl}

\usepackage[left=25mm,right=23mm,top=10mm,bottom=20mm]{geometry}
\usepackage{authblk}
\usepackage{graphicx}
\usepackage{booktabs}
\usepackage{subfig}
\usepackage{braket}
\usepackage{listings}
\usepackage{amsmath}
\usepackage{mathptmx} 
\usepackage{svg}
\usepackage[numbers,comma,sort&compress]{natbib} 
\usepackage[markcase=noupper]{scrlayer-scrpage}

\begin{document}
\title{Quantum computing in civil engineering: Potentials and Limitations} 
%
\author[1]{Joern Ploennigs}
\author[1]{Markus Berger}
\author[2]{Martin Mevissen}
\author[3]{Kay Smarsly}
\affil[1]{University of Rostock, Rostock, Germany\\
\{Joern.Ploennigs, Markus.Berger\}@uni-rostock.de}
\affil[2]{IBM Research Europe, Dublin, Ireland\\
martmevi@ie.ibm.com}
\affil[3]{Hamburg University of Technology, Hamburg, Germany\\
Kay.Smarsly@tuhh.de}


\setkomafont{title}{\normalfont\bfseries}
\setkomafont{author}{\normalfont\bfseries}
\setkomafont{section}{\normalfont\bfseries\Large}
\setkomafont{subsection}{\normalfont\bfseries}

%
%
\date{}
\maketitle              
\begin{addmargin}{-0.9cm}
\small
    \begin{abstract} 
      \textbf{Abstract}

      Quantum computing is a new computational paradigm with the potential to solve certain computationally challenging problems much faster than traditional approaches. Civil engineering encompasses many computationally challenging problems, which leads to the question of how well quantum computing is suitable for solving civil engineering problems and how much impact and implications to the field of civil engineering can be expected when deploying quantum computing for solving these problems. To address these questions, we will, in this paper, first introduce the fundamentals of quantum computing. Thereupon, we will analyze the problem classes to elucidate where quantum computing holds the potential to outperform traditional computers and, focusing on the limitations, where quantum computing is not considered the most suitable solution. Finally, we will review common complex computation use cases in civil engineering and evaluate the potential and the limitations of being improved by quantum computing. 

    \end{abstract}
\end{addmargin}
\section{Introduction}
Civil engineering is a complex area with many challenging problems in design, construction, and operation, which leads to the adaptation of many digital technologies from computer aided design, to construction robotics and machine learning, aiming to solve the underlying complex computational problems. The digitization of the industry is pushing the boundaries of what can be planned, analyzed, and optimized on current conventional computers, entailing a continuous need for new approaches.

Quantum computing (QC) is one of these new approaches. It may allow us to push some of the boundaries, as QC promises to compute numerous complex problems significantly faster than conventional computers. QC is explored in different industries \cite{bayerstadler2021industry,egger2020quantum}, but has not yet gotten much attention in civil engineering. The absence in civil engineering raises questions about the suitability of QC for addressing the computational challenges in civil engineering and the extent of its anticipated impact on the field. In this paper, we will review promising QC methods that may be be applied to current problems in civil engineering.

Therefore, we will first introduce the fundamental principles of quantum computing. We will delve into an analysis of scenarios where QC may demonstrate an advantage over traditional computers and identify areas where its application may be far in the future. With this context in mind, we will assess prevalent complex computational problems within civil engineering, determining problem classes that qualify for enhancement through QC. 




\section{Introduction to quantum computing}
\subsection{Fundamentals}
Traditional computers are based on processing and storing binary data in the form of bits that are either 0 or 1. A modern CPU is specialized on processing the binary data, usually in blocks of 64 bits, by applying logical operators (AND, OR, XOR, NOT), numerical operators (addition, substraction, multiplication and division), or specialized operators (encoding, encryption, etc.). Well established means have been proposed to represent any kind of data in binary code, such as numbers, texts, to images, and videos. The main assumptions behind binary encoding and processing are that (i) data is encodeable in binary 0 or 1 representation and (ii) all operators are strictly deterministic.

Quantum computing, by contrast, does not follow either of these assumptions, but uses a model of computation centered on \textit{quantum bits} or \textit{qubits} \cite{preskill2023quantum}. 
First, a qubit can be in a complex linear combination of both basis states $\ket{0}$ and $\ket{1}$, thanks to the principles of quantum \textit{superposition}, which allows a single qubit to represent multiple states in a Hilbert space. 

Secondly, quantum computers use the quantum mechanical effect of \emph{entanglement}, where the state of one qubit becomes linked or correlated with the state of another qubit, even when separated by large distances. With those entangled qubits it is possible to create quantum gates that implement specific quantum operations as exemplified in the next section. 

Third, due to the superposition principle (famously known as the Schrödinger's cat paradox), a qubit will collapse from its superposition state once measured and is identifiable only as one of the binary ground states. So, instead of being able to read the complex state between $\ket{0}$ and $\ket{1}$ directly, it is only possible to randomly sample binary states as either 0 or 1. Therefore, the qubit setup, computation and measurement operation needs to be repeated many times to retrieve a probably distribution representing the quantum state. 

\begin{figure}[t!]
	\centering
		\includegraphics[width=0.5\columnwidth]{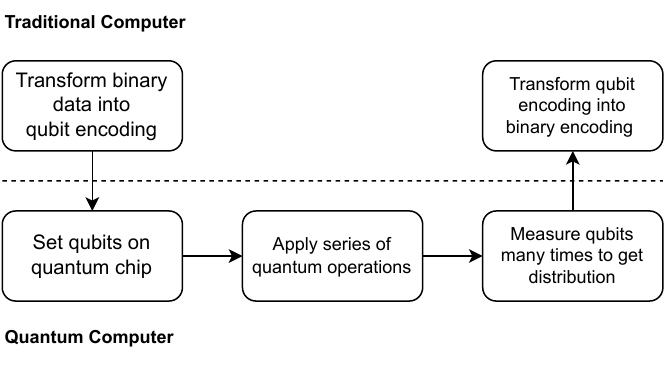}
		\caption{Process for combining traditional computers with quantum computers.}
	\label{fig:quantum_process}
\end{figure}

Fig.~\ref{fig:quantum_process} summarizes the process of loading data from a traditional computer into a quantum computer, applying quantum algorithms, and subsequently iterating through result retrieval. 
A number of architectures are currently being explored to realize universal gate-based quantum computers at scale. These include superconducting quantum computing, trapped ion systems, linear optical quantum computing and others. For example, IBM has published a development roadmap for its superconducting quantum computing systems for the next 10 years \cite{gambetta2023ibm}.

\subsection{Introduction to quantum circuits}
In this section we want to discuss exemplary the universal quantum gates that form quantum circuits to illustrate the complexity of quantum computation. 
To visualize the states of one qubit, we use the Bloch sphere model \cite{wiseman1993interpretation}.
On it the states of $\ket{0}$ or $\ket{1}$ are positioned at opposite poles. Points off the poles represents a superposition of qubit states. 

\begin{figure}[t!]
	\centering
		\includegraphics[width=0.8\columnwidth]{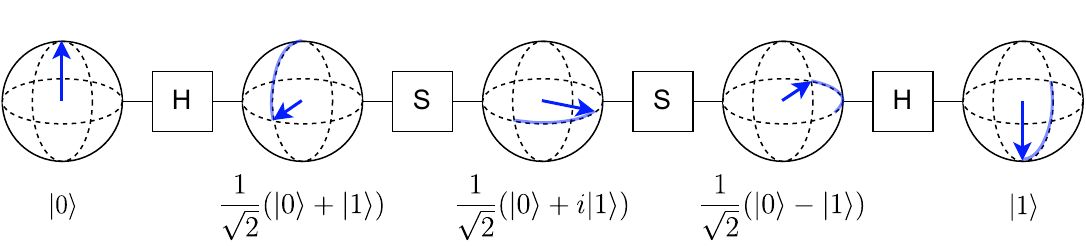}
		\caption{A quantum circuit with a series of Hadamard gates (H) and Phase gate (S) to move a qubit from a pure $\ket{0}$ to a pure $\ket{1}$ state shown as vector on a Bloch sphere}
	\label{fig:quantum_circuit}
\end{figure}

Quantum gates alter the states of the qubits on the surface of this sphere. 
Notable among these are as one-qubit gates the Hadamard gate (H), which initiates superposition of the initial states, and the Phase gate (S), which rotates the qubit state on the sphere through complex space \cite{nielsen2010quantum}. With these gates we can build a quantum circuit to negate a single qubit from $\ket{0}$ to $\ket{1}$ as shown in Fig.~\ref{fig:quantum_circuit}.

\begin{figure}[t!]
	\centering
		\includegraphics[width=0.45\columnwidth]{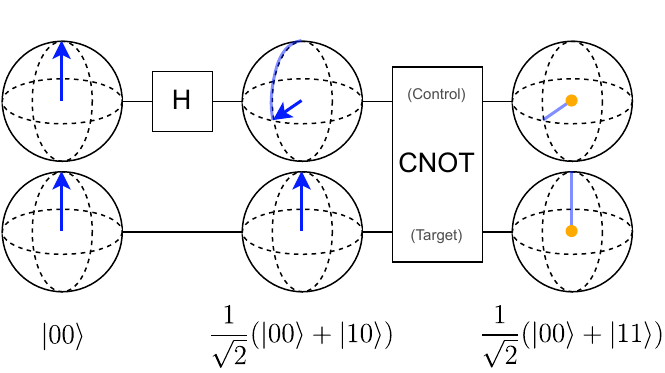}
		\caption{A quantum circuit with two qubits. The Hadamard gate (H) places the control qubit of the CNOT gate into a superposition state, resulting in the qubits becoming fully entangled. 
  }
	\label{fig:quantum_circuit_entangled}
\end{figure}

Another essential gate is the CNOT (Controlled-NOT-Gate) gate that operates on two entangled qubits. It allows to manipulate the state of one qubit based on the state of a second qubit. Specifically, the value of the second qubit (target) is either retained ($\ket{00} \to \ket{00}$; $\ket{01} \to \ket{01}$) or negated ($\ket{10} \to \ket{11}$;  $\ket{11} \to \ket{10}$) depending on whether the first qubit (control) is $\ket{0}$ or $\ket{1}$, respectively. Fig.~\ref{fig:quantum_circuit_entangled} shows an example. 


The CNOT gate and all one-qubit gates like the Hadamard and Phase gate form the set of universal gates that allow us to implement all other quantum gates and quantum algorithms. The resulting state of the target qubits at the end of the operation is measured multiple times, as discussed before. 

\subsection{Challenges}
There is a lot of ongoing fundamental research in quantum computing, and many challenges exist that limit its application at scale in practice. In this section, we aim to address some of the most prominent challenges, while dispelling common misconceptions about the current readiness of QC.


\textbf{Error handling:} 
Compared to classical computers, there are many sources of interference in quantum computing hardware, also called noise. This noise constrains the number of operational qubits. Current quantum computers are, therefore, \emph{noisy quantum computers}~\cite{preskill2018quantum}. As a result, one of the most prominent challenges in scaling up quantum computers is handling these errors affecting the actual quantum states and qubits. Broadly, we distinguish two approaches Error mitigation and Error correction \cite{cai2023quantum}.

\textit{Error mitigation} uses the outputs of quantum circuits to reduce the effect of noise in estimating expectation values, and is an approach to improve the performance of current noisy quantum computer in the short term. Two state-of-the-art techniques for error mitigation are Probabilistic Error Cancellation \cite{van2023probabilistic} and Zero-Noise Extrapolation \cite{he2020zero}. Those approaches estimate the noise and introduce additional computational overhead with limited scalability.


The long-term research targets Fault-Tolerant universal Quantum Computers (FTQC) by using error correction. \textit{Error correction}  is based on built-in redundancies, that is, on encoding each logical qubit through multiple redundant physical qubits \cite{postler2022demonstration}, and running error correction code for detecting and correcting errors. The redundancy, while critical for error correction, poses a significant challenge to the potential scalability of FTQC, as solving practical problems may require millions of qubits \cite{chatterjee2021semiconductor}. Thus, it is expected to take a number of years before realising first FTQC. Therefore, both approaches feature a trade-off between the additional overheads introduced by running the error mitigation or correction method on the one hand and the reduction in noise achieved on the other.

\textbf{Quantum Compilation:} Over the past few decades, we have become accustomed to the convenience of enhancing our computational performance by either upgrading hardware components or adding more machines due to standardized hardware, like the IBM PC, and operating systems, such as Linux, Windows, or iOS. This familiarity has led to an expectation that transitioning to quantum computing should be equally straightforward, allowing us to run our existing applications on quantum computers for vastly improved performance.

As shown in the introduction, quantum computers operate on fundamentally different principles than traditional computers, in terms of problem representation and solution methods. Consequently, we cannot simply cross-compile existing code to be run on a quantum computer in the same manner as we would do with a new CPU. Instead, each problem must be specifically adapted and implemented for quantum computation. Special programming languages for QC, such as IBM Qiskit, have been developed over the past few years \cite{heim2020quantum}. 
Nevertheless, the development of efficient quantum compilers remains a significant challenges, with many approaches under investigation how to optimization exact or approximate quantum compilation \cite{mariella2023doubly, madden2022best}. 
As a result, quantum computing programming languages, while resembling modern high-level programming languages, often operate at a very low level, close to the quantum gates themselves, and require a profound understanding of the underlying quantum computing models. Thankfully, experts develop problem specific frameworks, e.\,g. for quantum chemistry, machine learning or optimization problems that lower the entry barrier to utilize quantum computers for those use cases.

The development of GPU computing offers some parallels that can provide valuable insights. While modern PCs are equipped with GPUs boasting hundreds, if not thousands, of cores—far more than traditional CPUs—GPUs have not replaced CPUs because many computational problems in computing are not  \emph{embarrassingly parallelizable} and, thus, well-suited for GPUs. Instead, most problems are highly sequential, allowing the faster CPUs to solve them much more efficiently. Parallelizable problems also need to be specially implemented for the GPU in dedicated programming languages or frameworks. For instance, NVIDIA introduced CUDA already in 2007 \cite{kirk2007nvidia} with support for many programming languages. Yet, its applications in civil engineering software remain limited \cite{simpson2023challenges}. Instead, it took several years until the completely different area of neural networks took advantage of it to scale deep learning, which now has disruptive impact also on civil engineering.

\textbf{Quantum advantage:} Most existing software runs efficiently on traditional computers, both from a computational and from an energy perspective \cite{jaschke2023quantum}. For certain specific problem classes quantum computers have the potential to outperform classical computers in the near term. Here, we are speaking of \emph{quantum advantage}, when we are able to demonstrate that the quantum computer can outperform classical computers and find problem instances where this speedup is useful. 
Early quantum algorithms, such as Deutsch--Jozsa \cite{deutsch1985quantum}, Shor \cite{shor1994algorithms}, and Grover \cite{grover1996fast}, were ground-setting for this research area, but, have limited direct applicability on their own. 
The current focus is on finding problems where noisy quantum computers may provide computational advantage over classical computers considering that most algorithms require error correction, for example the widely re-used Shor algorithm for integer factorization. Indeed, there have been recent results, demonstrating the utility of quantum computing for simulating the time evolution of 2D transverse-field Ising models \cite{kim2023evidence}. There is a lot of ongoing research at the moment, in order to find further challenging problems, where noisy quantum computers---along with techniques like error mitigation---are demonstrated to be a useful tool.



\section{Use cases}
\subsection{Use cases of quantum computing}

In this section, we will explore use cases that are expected to exhibit a quantum advantage in the near future and share similarities with problems encountered in civil engineering. The use cases encompass: (i) \emph{simulating} natural physical processes, 
(ii) \emph{mathematical and machine learning algorithms} to process data with complex structure, and (iii) solving \emph{optimization} problems as well as search.

\textbf{Simulating} nature represents one significant application area for quantum computers, particularly for systems operating (at or close to) the quantum level, such as simulating Hamiltonian dynamics or preparation of ground states. There is an increasing body of work that studies the properties of many body quantum systems with the help of quantum computers, e.\,g. \cite{keenan2023evidence}. Current exploration in the field includes applications in chemistry, such as the study of molecular structures \cite{outeiral2021prospects} or chemical reactions \cite{cao2019quantum}. Quantum computers facilitate the investigation of complex materials, including superconductors and novel compounds, allowing for more efficient comprehension of the properties and potential applications compared to classical methods. Leveraging the precision in simulating quantum mechanics, quantum computers are well-suited for providing more accurate predictions of molecular behavior, which is promising for future developments in drug discovery, materials design, and catalyst development \cite{daley2022practical}.
A particularly encouraging recent result has been the demonstration of the utility of quantum computing for simulating the time evolution of 2D transverse-field Ising models  \cite{kim2023evidence}.

Certain \textbf{mathematical and Machine Learning} problems and processing data with complex structure are expected to benefit from approaches based on quantum computing. In particular, demonstrating exponential speed-up is expected for quantum machine learning methods. 
While theoretic development of quantum machine learning algorithms goes back several decades, there has been a growing body of work on quantum machine learning approaches explored on current noisy quantum computers. 

A promising class of quantum machine learning approaches are quantum kernel methods for classification problems, one of the most fundamental problems in machine learning. In a recent result, it was shown that there are specific classification tasks where a quantum kernel method with only classical access to data provides an exponential speed-up over classical machine learning algorithms \cite{liu2021rigorous}. It is an ongoing direction of research to develop supervised quantum machine learning algorithms for broader sets classification tasks.

Another notable group of quantum algorithms with potential applications in machine learning includes quantum phase estimation (QPE) \cite{dorner2009optimal} and the underlying quantum Fourier transform (QFT) \cite{hales2000improved}, which is similar to the classical discrete Fourier transform and performs a Fourier transformation of the amplitudes but in $O(n^3)$ whereby the best known classical algorithm requires $2^{O(n^{1/3})}$ time. The QFT finds widespread usage in various other algorithms, including Shor’s algorithm for integer factoring \cite{shor1994algorithms}, the Harrow-Hassidim-Lloyd (HHL) algorithm for solving linear equations \cite{harrow2009quantum}, and quantum gradient estimation \cite{jordan2005fast}. One application area of Shor's algorithm is cryptography, where the algorithm enables  efficient factorization of integers to find the prime factors.


The advancements made in solving specific mathematical matrix problems using QAOA have found applications in enhancing traditional machine learning models, including linear regression \cite{date2021adiabatic}, clustering \cite{gao2022quantum, horn2001algorithm}, reinforcement learning \cite{dong2008quantum, lamata2017basic}, support vector machines \cite{havlivcek2019supervised}, and active learning \cite{paparo2014quantum}. Additionally, there is ongoing exploration of quantum algorithms for neural networks \cite{kerenidis2019quantum, schuld2014quest}. Furthermore, quantum computing has been investigated for its potential to improve classical machine learning algorithms, particularly in the context of handling tensor and dot products in higher dimensions, thereby reducing computation time \cite{dasari2020solving, Fastovets2019}. In all these cases, quantum computing addresses specific aspects of traditional algorithms.

A sub-category of the aforementioned problems includes classical data analysis and statistics problems, which can be advanced through machine learning algorithms and, also, by more classically adapted methods, such as quantum principal component analysis \cite{lloyd2014quantum} or quantum clustering \cite{aimeur2007quantum}. 
Again, quantum advantage has not be shown to be practical for any of these models, yet, and achieving it on current noisy quantum computers is challenging due to the necessity of conducting numerous measurement runs and the stochastic nature of the objective function.

\textbf{Optimization} is another class of problems where quantum computing may be demonstrated to be useful, even if no exponential speed-up over classical computing is expected. However, various problems may also derive benefits from quantum computing \cite{abbas2023quantum}. 
Several gate-based optimization algorithms have been developed, involving the simulation of the evolution of a system through a sequence of quantum operators. Two common methods include the variational quantum eigensolver (VQE) \cite{peruzzo2014variational} and the quantum approximate  optimization algorithm (QAOA). Both methods are meta-heuristics for solving combinatorial optimization problems that can leverage gate-based quantum computers and potentially outperform purely classical heuristic algorithms \cite{fuchs2022constraint}. The approaches mentioned above find application in logistical problems \cite{ajagekar2020quantum, wang2023opportunities}, such as optimizing traffic networks to reduce congestion and enhance efficiency \cite{yarkoni2020quantum} or formulating and solving routing problems \cite{harwood2021formulating}. Similarly, the methods are employed for optimizing energy distribution networks \cite{ajagekar2019quantum}. A common feature among these applications is the utilization of a graph representation \cite{goodrich2018optimizing}, where the problems are modeled as a Max-Cut problem.

Finding an exact solution to the Max-Cut problem is known to be NP-hard \cite{karp2010reducibility}. The objective of the Max-Cut problem is to divide the set of graph vertices into two subsets in a way that the sum of the weights of the edges connecting one subset to the other is maximized. Solving Max-Cut with QAOA has seen some focus since it can be formulated as an unconstrained quadratic unconstrained binary optimization (QUBO) problem and has wide number of applications. The approach with QAOA is based on the following mapping:
\begin{equation}
    \arg \min_x \sum_{i,j} w_{i,j} x_i x_j, \qquad \xrightarrow[\text{to QAOA}]{\text{mapped}} \qquad H = \sum_{i,j} w_{i,j} \sigma_i \sigma_j.
\end{equation}
where $w_{i,j}$ are the weights for the binary assignment $x_i$ and $x_j$. To solve the problem using QAOA, we construct the cost Hamiltonian $H$ by mapping the binary assignment variables $x_k$ onto the eigenvalues of the Pauli Z operator $\sigma_k$. By employing the mapping, we obtain a representation on the right that closely resembles the original problem. It is expected, albeit unproven, that a quantum computer can solve this problem more efficiently than a classical one \cite{shaydulin2019evaluating}.


\subsection{Use cases in civil engineering}
Based on this discussion of use cases for quantum computers we can derive some insights on which use cases in civil engineering may be most appropriate to benefit from quantum computers. We derive the mapping shown in Fig. 4 with good (bold), some (dotted) and limited (none) applicability as explained below.


\begin{figure}[t!]
	\centering
		\includegraphics[width=0.6\columnwidth]{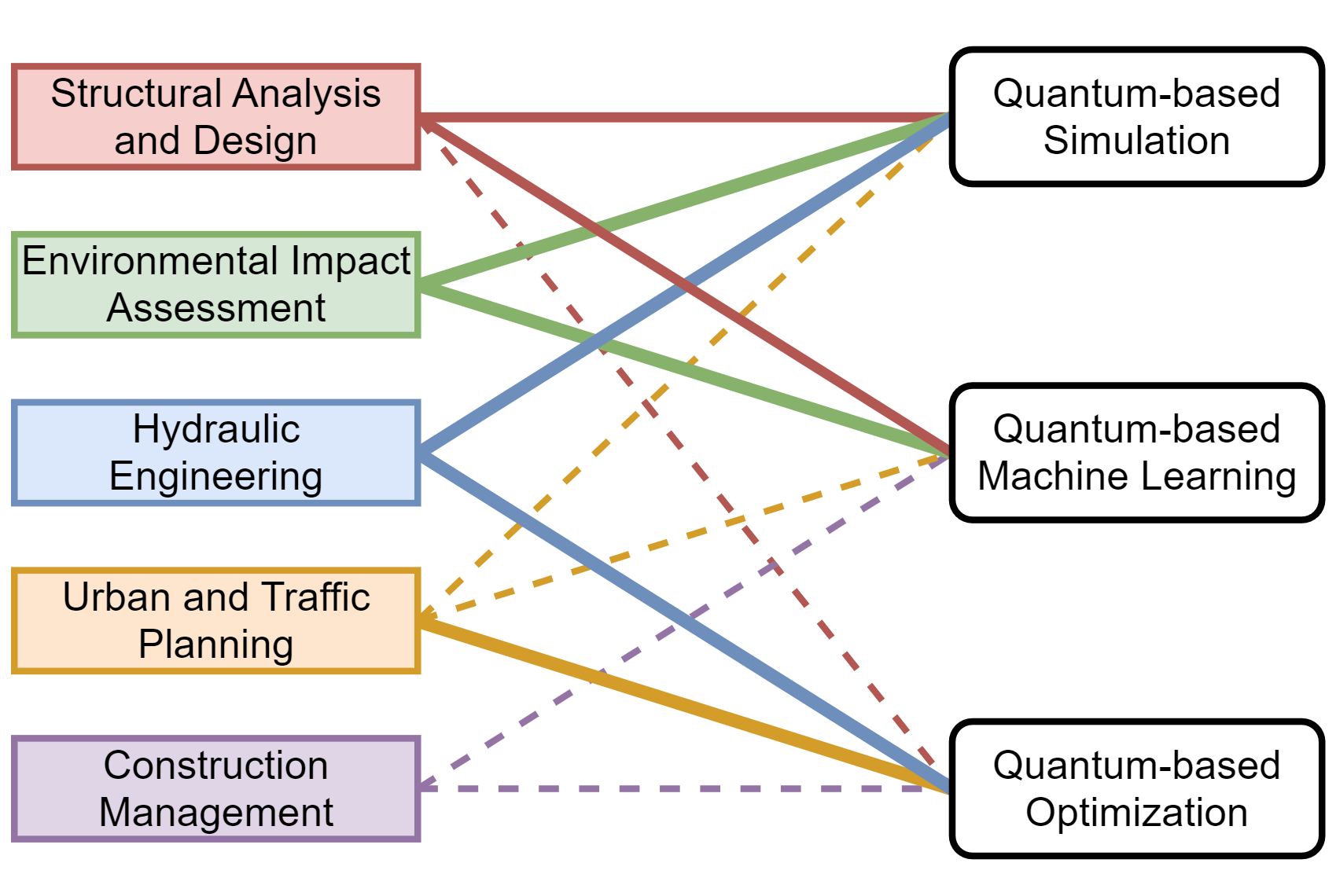}
		\caption{An overview of engineering disciplines associated with the potential applicability of quantum computing. Lines highlight good (bold), some (dotted), or limited (none) applicability.}
	\label{fig:algorithms_overview}
\end{figure}

\textbf{Simulations} are a common approach in civil engineering to evaluate design variants or to understand real-world phenomena. Some of those approaches may benefit from quantum computing.

\textit{Urban and traffic planning} is using simulations specifically for modeling people and traffic flows and congestion within transportation networks, encompassing interactions among vehicles, infrastructure, and traffic signals. Traffic simulations rely on computational models that account for a multitude of variables, ranging from vehicle behavior and road conditions to traffic patterns \cite{treiber2013traffic}. These simulation models are less suited for quantum.

\textit{Structural analysis and design} is commonly employing either rigid body models or finite element analysis (FEA) to simulate and analyze the behavior of structures under various conditions, including stress, heat transfer, and fluid flow. FEA, in particular, demands substantial computational resources as it involves dividing a structure into numerous small interacting elements to approximate the overall system behavior \cite{zienkiewicz2005finite}. \textit{Geotechnical engineering} problems are another application area for FEA. Here, engineers simulate soil-structure interactions, slope stability analysis, and foundation design \cite{das2021principles}. \textit{Seismic analysis and design} combines aspects of both structural and geotechnical engineering and focuses on designing structures capable of withstanding seismic forces, entailing complex analyses of how structures respond to earthquakes \cite{humar2012dynamics}. 

\textit{Hydraulic engineering} often employs computational fluid dynamics (CFD) models to analyze the behavior of rivers, dams, coastal areas, and other hydraulic structures. The models, similar to FEA, break down problems into small interacting fluid (or gas) elements, but emphasize the solution of intricate fluid dynamics equations such as turbulence modeling, sediment transport, and interactions with structures \cite{anderson2020computational}.

\textit{Environmental Impact Assessment} is looking into simulating the environmental impact of constructions often applying a combination of CFD or FEA methods \cite{HARISH20161272}.


The FEA and CFD problems fundamentally involve simulating the underlying physics in the form of higher-order differential equations, without relying on specific quantum mechanical effects for which quantum computers are well-suited. Instead, quantum algorithms that enhance the efficiency of solving differential equations, such as those breaking them down into systems of linear equations solvable with algorithms, as the HHL algorithm \cite{harrow2009quantum}, are explored \cite{berry2014high, liu2021efficient}.

A few applications to civil engineering problems can be found. Ajagekar and You~\cite{ajagekar2019quantum} show a heat exchanger network synthesis problem, where multiple cooling units need to be balanced. Quantum approaches are utilized to simulate the behaviour and to identify the optimal solution.

Various \textbf{Mathematical and Machine Learning} problems exist in civil engineering, primarily in operation scenarios, but also in design and construction. 

\textit{Structural analysis and design} is using methods for \textit{structural health monitoring} to record and analyze sensor data for assessing structural conditions of infrastructure, including the detection of defects or damage, aiming to support predictive maintenance and life-cycle management \cite{flah2021machine,Smarsly16}. \textit{Predictive maintenance} is a related topic that focuses on analyzing system performance to predict maintenance actions \cite{cheng2020data}, applicable not only to structures but to various types of assets.

\textit{Environmental Impact Assessment} is another field using prediction models to analyze performance measures in general for a system like their energy consumption \cite{ahmed2011mining}. \textit{Performance-based design} can conceptually use the models to evaluate different design options under various conditions or environmental impacts. If the prediction models are trained from simulation models, we speak of \textit{surrogate models} \cite{sun2021machine}.

\textit{Hydraulic and Energy Engineering} is commonly targeting \textit{demand prediction} of consumers in energy or water grids, or the traffic flow of people and cars in \textit{Urban and Traffic Planning}. The objectives are similar to the simulation scenarios discussed earlier, but rely on data-driven models instead of simulation models \cite{lassoued2017hidden}. 


The specific machine learning models employed depend on the use case and data type. Typically, the data is recorded through sensors and stored in the form of time series. Regression models, such as linear regression or support vector machines, are often sufficient for prediction \cite{pathak2018forecasting}. Quantum computing variants exist for these models, as discussed earlier \cite{date2021adiabatic, havlivcek2019supervised}, but their necessity is debatable, as time series problems can be decomposed and trained individually \cite{patel2023ai}. For problems with a graph structure, such as energy, water or traffic networks, graph neural networks may be more appropriate \cite{ba2024}, although more performant quantum variants are currently lacking. In cases involving image or video data sets, deep learning models are commonly used, as the models tend to benefit more from GPU processing than quantum computation do \cite{date2020classical, combarro2023practical}.

\textbf{Optimization} in civil engineering involves finding the best arrangement, combination, or configuration of elements within a discrete set of choices, considering various constraints and objectives. 

\textit{Construction Management} deals with efficiently assigning resources (e.\,g. labor, equipment, materials) in allocation and scheduling problems and schedule construction activities\cite{hinze2004construction}. These problems can often be modelled as a QUBO \cite{kochenberger2006unified}. 

\textit{Urban and traffic planning} addresses the design of optimal traffic networks \cite{toth2014vehicle}, which is an extension of the logistic routing problem discussed earlier \cite{ajagekar2020quantum, wang2023opportunities}. However, there has not been a discussion yet on the evaluation of the impact of the algorithms on network design. Conceptually related is the \textit{site location selection} problem, which refers to determining the most suitable locations for infrastructure facilities, such as schools, hospitals, waste management sites, or transportation hubs. The determinations take into account factors, such as accessibility, environmental impact, population distribution, and cost \cite{Donncha23}. The problems can often be modeled as coverage problems \cite{farahani2012covering}, and thus we may be able to map the problems onto a QUBO problem, which, in turn, might be solved by QAOA. However, in practice, the number of constraints and weights is often higher, as demonstrated by Farahani et al.~\cite{farahani2012covering}.

\textit{Hydraulic engineering} and distribution networks planning problems are prevalent in civil engineering, spanning various forms from power to fresh and waste water networks. Early work explores the use of quantum computers for optimizing energy distribution networks \cite{ajagekar2019quantum}. Similar approaches may apply to other network types by mapping the problems to graph Max-Cut problems.

\textit{Environmental Impact Assessment} is usually more looking in using optimization for improving energy efficiency. Here the optimization approaches are either control problems or scheduling problems.

\textit{Structural analysis and design} is optimizing the arrangement and sizing of structural components (beams, columns) that ensure structural integrity while minimizing material usage\cite{belegundu2019optimization,mei2021structural}. Optimization often requires solving complex mathematical problems, such as nonlinear programming or genetic algorithms. Considering the problem as a rigid body problem with a graph representation may allow to map it to QADA. 
For example, Wang et al.~\cite{wang2023opportunities} discuss an application for a topology optimization problem called the Messerschmitt-B\"olkow-Blohm (MBB) problem \cite{bendsoe2003topology} to find the stiffest design of a desired volume fraction. Traditionally, such design problems were solved by engineers through experience or trial and error.

\section{Summary and Conclusions}

In this paper, we have discussed the current state of quantum computing and potential applications in civil engineering. We first provided an overview of the fundamental concepts of quantum computing, highlighting its distinctions from traditional binary computing. Subsequently, we have addressed the current challenges associated with scaling quantum computing, with a specific focus on algorithms and problem classes that can be studied with current, noisy quantum computers. We then have reviewed the areas where quantum computing has the potential to outperform traditional computing, namely in simulating quantum mechanics, quantum machine learning approaches, and classes of optimization problems. For each area, we have also introduced the relevant use cases in civil engineering and have discussed the potential of applying QC within them. The preliminary conclusions are summarized as follows:
\begin{description}
    \item[Simulation:] Among the most computationally demanding tasks are FEA and CFD problems, where quantum computing holds the potential to enhance performance in solving differential equations. Here, the discretization into linear equation systems is an important step.
    \item[Machine Learning:] Quantum computing shows the largest promise in the speed-up that may be achieved through quantum kernel methods for a variety of classification tasks. In addition, advancements in machine learning involve algorithmic steps that may have the potential to perform better on a quantum computer. To benefit from these future possibilities, it is recommended to keep investigating the applicability of these classification models in civil engineering.
    \item[Optimization:] Quantum computing has already been explored for a number of optimization problems arising across different application domains. There is a lot of ongoing research in the speed up that may be achieved for particular algorithms. In civil engineering this means to identify relevant problem-mappings to solutions like Max-Cut or QUBO.
 \end{description}

Quantum computing has the potential to catalyze a revolutionary paradigm shift similar to the transformative impact observed with GPU computing, currently reshaping the landscape of science through the empowerment of deep learning. The innovations in quantum computing will not arise primarily from accelerating existing code; instead, innovations are expected to derive from the ability to address an entirely new set of challenges using specialized software applicable across a broad spectrum of domains.

%
%
%
\begingroup
    \small
    \setlength{\bibsep}{0pt}
    \bibliography{references}
\endgroup
\end{document}